# Operation of a Continuous Flow Liquid Helium Magnetic Microscopy Cryostat as a Closed Cycle System

K. Barr[1,a)], T. Cookson[2,3] and K. G. Lagoudakis[1,a)]

**AFFILIATIONS**

[1]Department of Physics, University of Strathclyde, Glasgow G4 0NG, United Kingdom
[2]Department of Physics and Astronomy, University of Southampton, Southampton SO17 1BJ, United Kingdom
[3]Centre for Photonics and Quantum Materials, Skolkovo institute of science and technology, Skolkovo, Moscow 121205, Russian Federation

[a)]Author to whom correspondence should be addressed: k.barr@strath.ac.uk and k.lagoudakis@strath.ac.uk

**ABSTRACT**

We demonstrate successful operation of a continuous flow liquid helium magnetic cryostat (Oxford Instruments, Microstat MO) in closed cycle operation using a modular cryocooling system (ColdEdge Technologies, Stinger). For the system operation, we have developed a custom gas handling manifold and we show that despite the lower cooling power of the cryocooler with respect to the nominal cryostat cooling power requirements, the magnetic cryostat can be operated in a stable manner. We provide the design of the gas handling manifold, and a detailed analysis of the system performance in terms of cooling times, magnetic field ramping rates and vibrations at the sample. Base temperature can be reached within 10 hours while the superconducting magnet can be energized at a ramping rate of 0.5 T/min. Vibrations are measured interferometrically and show amplitudes with a root mean square on the order of 5 nm permitting the use of the system for sensitive magnetic microscopy experiments.

## I. INTRODUCTION

With increasing scarcity[1-3] and continuous inflation of Helium prices[4], several experimental groups with setups that up until recently used continuous flow liquid Helium (LHe) cryostats for measurements at cryogenic temperatures, find it increasingly difficult to maintain operation of the setups. In an effort to circumvent the high operational costs and the frequent supply shortages, some research groups have invested in a one-off manner, in new closed cycle cryorefrigeration systems[5], leaving perfectly operational continuous flow cryostats unused.
Recently, many companies have been developing closed cycle cryocoolers that are compatible with existing continuous flow cryostats[6-8], enabling their operation as closed cycle systems. Small cryostats for cryo-microscopy require relatively low cooling powers and therefore can be operated with commercially available cryocoolers right away. However, larger continuous flow cryostats like those for magnetic cryo-microscopy require higher cooling powers[9] and at first sight, might seem to have cooling power requirements that exceed the power of commercially available cryocoolers. In this letter, we report our investigation on the performance of such a continuous flow magnetic microscopy cryostat (Oxford Instruments *Microstat MO*) when it is cooled by a commercially available cryocooler (ColdEdge Technologies *Stinger*) in terms of cooling behavior, superconducting magnet operation and most importantly sample vibrations.
We demonstrate that despite the cooling power mismatch between the available cooling power of the cryocooler and the cryostat needs, the cryostat can be successfully cooled down in approximately 10 hrs, the magnet can be ramped up to 5 T at a rate of 0.5 T/min, and that the sample is maintained at a steady temperature of about 8 K with approximately 5 nm root mean square (RMS) amplitude vibrations.

Section II describes the complete system along with the custom helium handling manifold that we developed for the operation of the cryostat. Section III then describes the heating and cooling processes from room temperature to baseline cryogenic temperatures. Section IV describes a method for determining air contamination within the system. Section V consists of vibrational analysis in the y- and z-axis of the sample holder in the sample well and suggests methods to minimize external vibrational sources that can affect the system performance. Section VI shows the performance of the superconducting magnet, reaching a peak magnetic field of 5 T within ten minutes, while section VII describes the quench dynamics of the superconducting magnet and the recovery of the system.

## II. SYSTEM DESCRIPTION

The MicrostatMO is a continuous flow (open cycle), cryostat with a superconducting magnet commonly used for cryo-microscopy experiments that require the presence of high (up to 5T) magnetic fields while operating at a wide range of temperatures (6 - 310 K). Microscopy is possible as the sample is kept ~8 mm from the sample chamber window, making it optically accessible using long working distance microscope objectives[10].

The cryostat has a nominal LHe consumption of 2 l/h which, using the latent heat of vaporization of helium, can be translated to a cooling power requirement of approximately 1.44 W.

The cryostat is typically supplied with LHe by means of a low loss coaxial LHe transfer tube and features two helium return paths, one along the coaxial transfer tube which we call the main return path, and one return path through the high current lead tube, which we call the side arm return path, as shown in Fig. 1 (a). The Stinger is supplied with a similar format transfer tube insert, so it can be directly connected to the cryostat without requiring any modifications. The complete setup for the operation of the cryostat as a closed cycle system, is shown in Fig. 1 (b).

Operation of the superconducting magnet requires a fine balance between the helium gas flow rates of the main and side arm return paths. As the off-the-shelf helium handling manifold that comes with the Stinger does not include flowmeters, we developed a custom made helium handling manifold combining all the necessary features for the operation of the Stinger and the cryostat in one unit. In particular we introduced two Helium glass variable area flowmeters (Platon NGXV S312 with glass scales FGTF-2AHD 2 - 28 l/min and FGTF-2BHD 0.5 - 9 l/min calibrated for helium gas) as shown in Fig. 2. As the flowmeters are equipped with needle valves that enable precise flow control, we incorporated bypass relief valves that eliminate accidental over-pressurization of the helium flow circuit in the cryostat, if a user shuts both valves. An additional pressure relief valve that vents to the room is connected to the helium return side and ensures that the overall pressure in the helium return circuit never exceeds the limit value of 20 psi, as shown in the diagram of Fig 2 (a).

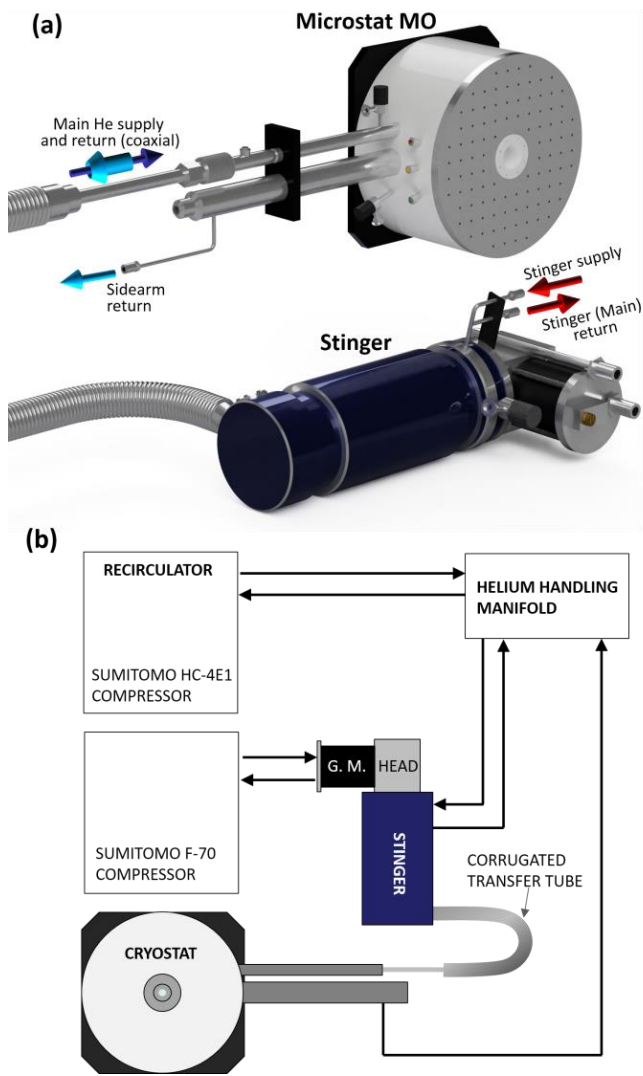

**FIG. 1.** (a) Rendering of the cryostat (top) with the Stinger cryocooler (bottom) attached via the corrugated transfer line tube. The main helium supply and return paths for the cryostat are housed within the corrugated helium transfer tube which is only partially shown for compactness. The arrows depict the helium flow in and out of the system with their color associated to the relative temperatures (blue: base T, teal: intermediate T, red: room T). The red, yellow and green dots on the cryostat casing are the electronic connectors for monitoring the magnet and controlling the sample temperature. (b) Schematic diagram showing the complete cryostat-cryocooler system along with the associated compressors. The recirculator is used in combination with the helium handling manifold (see Fig. 2) to circulate helium in and out of the Stinger and the cryostat. The Stinger is cooled by a Gifford-McMahon cold head that is driven by the F-70 compressor. All thick black arrows depict the flow direction of the helium in the flexible lines.

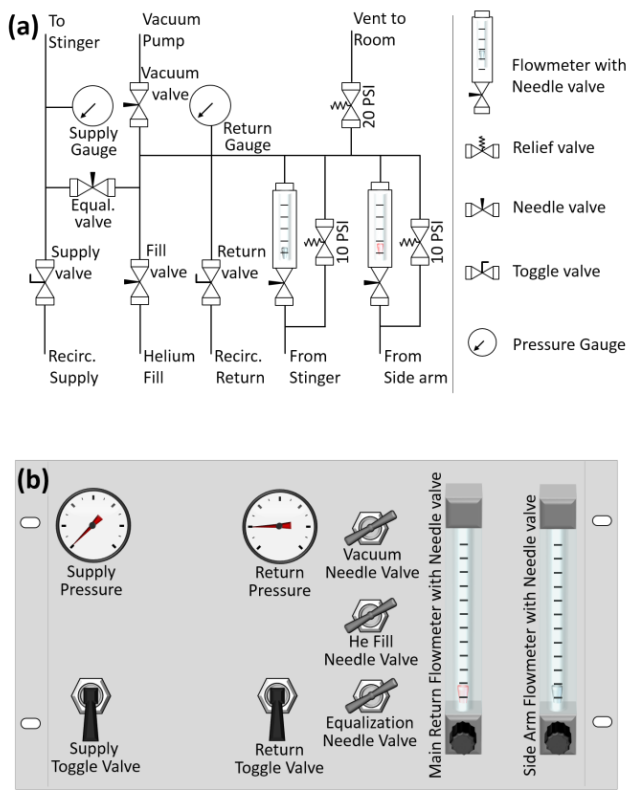

**FIG. 2**. Custom built helium handling manifold developed for interfacing the Oxford Microstat MO cryostat with the Coldedge Stinger cryocooler. (a) Internal connection diagram. To avoid overpressurization of the helium circuit of the cryostat, bypass relief valves have been connected in parallel to the flowmeters and a 20 psi relief valve venting to the room ensures that pressure is never above that value in the helium circuit of the cryostat. When opened, the equalization valve allows simultaneous pumping of both supply and return helium circuits. All lines that cross are connected. (b) Front panel design of the control manifold. The supply pressure gauge has a range 0 - 160 psi while the return pressure gauge has a range -30 - 60 psi.

To evacuate the system prior to operation, an equalization valve connects the supply and the return paths, and in combination with the vacuum valve allows pumping of both sides simultaneously for the elimination of air contamination from the helium circuit. All tubing within the manifold is seamless stainless steel while all fittings are double ferrule stainless steel. The manifold is housed in a 19" rack enclosure and the components are laid out as shown in Fig. 2 (b).

Helium gas is forced through the system using a helium recirculator which consists of a Sumitomo HC-4E compressor. The heart of the Stinger is a Sumitomo Gifford McMahon cryocooler (Model RDK-408D2 with 1.0 W cooling power at 4.2 K and is driven by a Sumitomo F-70 compressor) that provides the bulk of the helium gas cooling. Whilst the cryocooler provides the primary refrigeration of the helium flowing through the Stinger at approximately 110 psi, the helium is then directed into the flexible corrugated transfer tube that is connected to the cryostat. At the end of the transfer tube, the flow undergoes Joule-Thomson expansion from 110 psi to atmospheric pressure, leading to further cooling, achieving temperatures of 4 K or lower. Joule-Thomson expansion is a technique, which the cryocooler adopts, whereby a change of pressure in the system can allow helium expanding into the lower pressure region to be further cooled down. Typically, the Stinger generates 0.5 W of overall refrigeration capacity at 4.2 K, which has proven to be sufficient for the operation of the magnetic cryostat.

## III. CHARACTERIZATION OF TEMPERATURE PROFILES

A typical cryostat cool down process begins by evacuating the helium circuit in the manifold and the cryostat to high vacuum, to ensure that all air contamination is removed. This helps to prevent the system getting clogged with frozen air once base temperatures are reached. It has been found that a vacuum pressure of the order $\sim 1\cdot 10^{-3}$ mbar has proven effective. Opening the return and supply toggle valves on the manifold, fills the supply circuit with helium in the range of 100 - 110 psi. As the cryocooler cools the circulated helium to base temperature, liquefaction of helium gradually takes place first in the Stinger coldhead and later in the cryostat itself and the overall pressure drops by about 50 psi[10]. It is therefore important to initialize the system at room temperature with helium pressure in the range of 100 - 110 psi, to ensure that helium can still flow at the rates necessary for maintaining the system at base temperature. Fig. 3 (a), shows the temperature profiles for a typical cooling cycle as a function of time. We see that it takes slightly over one hour for the Stinger to reach a base temperature of approximately 6 K temperatures. When the system is at base temperature, typical helium flow rates through the magnet and sidearm are approximately 15 l/min and 6.5 l/min, respectively.

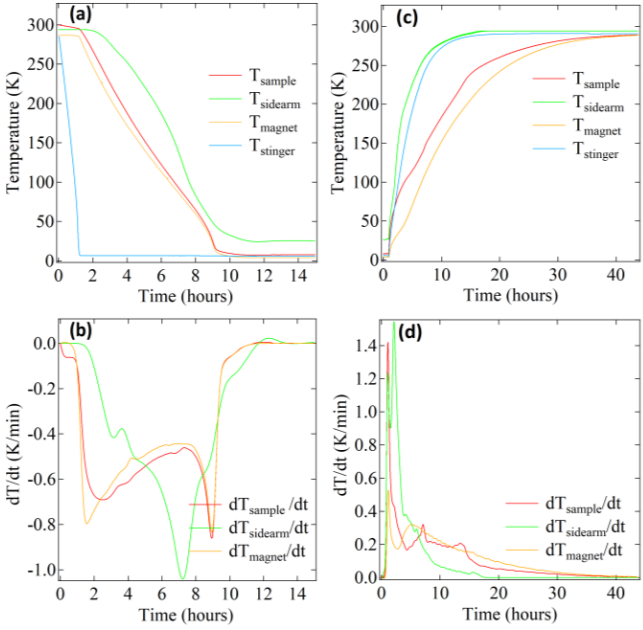

**FIG. 3**. Cool-down and warm-up temperature profiles for the system. (a) Temporal behavior of cryostat temperatures during cool-down. The Stinger reaches base temperature within ∼1 h, after which the cryostat starts to cool-down and reaches base temperature after ∼10 h. (b) Temperature gradients during cool-down for the sample, magnet and side arm. (c) Temporal behavior of cryostat temperatures and (d) temperature gradients during warm-up for the 3 cryostat components. Unless the cryostat vacuum is broken, the warm-up is slower than the cool-down by a fivefold taking ∼48 h overall.

The process of heating up is a considerably longer process. To warm up the cryostat we turn off the Stinger compressor and the helium supply toggle valve on the manifold, while we keep the recirculator running to absorb all the helium that quickly boils off thereafter. Once the helium flows drop to zero, we shut the return toggle valve and allow the system to warm up. Letting the system warm up naturally, takes about 48 h with rates that gradually slow down, as shown in Figs. 3 (c) and (d). This is due to the low thermal conductivity of the system if one maintains high vacuum in the cryostat casing. Should the need to heat up the system in less time be of importance, it is possible to break the vacuum in the cryostat to speed up the process.

## IV. SIGNS OF AIR CONTAMINATION

Any air contamination within the helium circuit will lead to many issues during a cooling cycle and often prevent the system from reaching or maintaining operational temperature. As such, it becomes highly beneficial to be able to identify such an issue at an early stage. As air freezes at LHe temperatures, it can cause blockages within the system. Fig. 4 at which point the rest of the system begins to follow. The magnet and sample holder cool at similar rates, with the magnet exhibiting an initial cooling rate of -0.8 K/min which gradually slows until it levels-off at around -0.4 K/min, as shown in Fig. 3 (b). When the magnet reaches 50 K, an increase in the rate of change occurs for a short period until the base temperatures are reached. Beyond this point, minor adjustments can be applied to the flow rates to both the magnet and sidearm until the desired temperatures are reached. To avoid quenching of the superconducting magnet during operation[10], the magnet should be kept under 5.5 K whereas the sidearm should be under 35 K. Once the system reaches steady state, fine tuning of the flow rates through the manifold flowmeter needle valves, settle the temperatures at around 4.15 K for the magnet and 29 K for the sidearm. While it is not a necessary requirement to reach such low temperatures for the magnet and the side arm, the system naturally falls to these compares the behavior of the magnet temperature under normal operation and in the presence of air contamination. In particular, Fig. 4 (a) compares the temperature profiles of the Stinger and the magnet under normal operation and in the presence of air contamination while Fig. 4 (b), shows the temperature gradient of the magnet for both cases. Under normal operation, a gradient of -0.8 K/min is reached when the Stinger reaches base temperature, about ~1.5 h after cooling initialization, followed by a slow deceleration of the cooling rate to ~ -0.4 K/min. In the presence of air contamination though, the magnet cooling rate reaches a maximum value of -0.3 K/min and quickly stabilizes to -0.1 K/min after that, highlighting the blockage because of the air contamination. This striking difference in the cooling rate of the magnet, should allow one to identify blockages caused by frozen air within the helium circuit early.

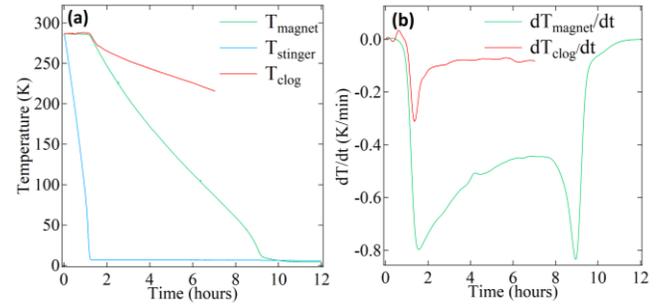

**FIG. 4**. (a) Temperature profiles of the magnet under normal operation (green) and in the presence of air contamination (red). The Stinger temperature is shown in blue for a visual comparison. (b) Temporal gradients of the magnet temperature throughout a cool-down cycle in the presence (red), and in the absence (green) of air contamination. The acquisition of data was interrupted after 7h for the air contaminated case.

### V. VIBRATIONS

One of the key drawbacks of closed cycle systems in general, is the issue of vibrations generated by the cryocoolers. In the system we investigate here, the coldhead on the Stinger and its compressor are considerably noisy, in the sense that they can and do introduce unwanted vibrations into the system. The closed cycle system introduces vibrations to the sample holder in numerous ways, where the dominant predicted channel is the mechanical coupling directly to the cryostat through the corrugated flexible helium transfer tube. What follows is an analysis of the vibrations introduced to the system by the cryocooler and steps taken to minimize their effects. Although several methods exist for the characterization of vibrations in cryogenic environments[11-15], here we employ an interferometric approach[15-17]. To reduce the acoustic noise picked up by the interferometer, we enclosed both the Stinger and its compressor into a custom casing surrounded by aluminum dibond panels that were padded by 25 mm thick, class 'O' acoustic foam. To further reduce any residual acoustic noise coming from the compressor, we placed the compressor and recirculator in an adjacent room. The Stinger which is the main component that is mechanically coupled to the cryostat, is required to be nearby the cryostat because of the finite length of the helium transfer tube, so in our setup it is placed right under the optical table. The use of acoustic foam in the enclosure of the Stinger head proved to significantly reduce the observed vibrations in the interferometer by approximately a factor of 2. The helium transfer tube that is connected directly to the cryostat was also clamped to the table using a thin layer of acoustic foam on the contact point in an effort to further damp vibrations. For the interferometric vibration measurements we used a frequency stabilized HeNe laser. For the measurement of the vibrations in the z-direction (along the optical axis of the cryostat) we built a Michelson interferometer, like the one depicted in Fig. 5 (a). The vibrations show up as intensity fluctuations on the photodiode signal and given the known wavelength of the laser and the geometry of the system, they can be directly converted to distance. In particular, to get from constructive to destructive interference it takes a λ/4 displacement of the sample holder because of the double pass of the laser beam to and from the sample holder. As the highest sensitivity only occurs for signals away from the maximum and minimum of the interference signal, we always performed the interferometric measurement when the photodiode signal was within the band of 30-70% of the peak to peak interference fringe amplitude. By removing slow signal variations[10] with (ν < 0.1Hz), one can estimate quantities such as the root mean square (RMS) and average deviation (ADEV) of the vibrations directly. Fig. 5 (b), shows the vibration data of a typical acquisition in the z-direction. The peak to peak amplitude is ∼30 nm, the RMS is found to be ∼5 nm whereas the ADEV is estimated to ∼4 nm. The vibrational spectra are extracted by taking the Fast Fourier Transform (FFT) of the time trace, as shown in Fig. 5 (c). Interestingly, the spectra do not feature peaks at 2 Hz which is the typical mechanical motion of the GM cryocooler in the Stinger but rather show distinct frequency peaks with two dominant frequencies of vibration at 101 Hz and 300 Hz.

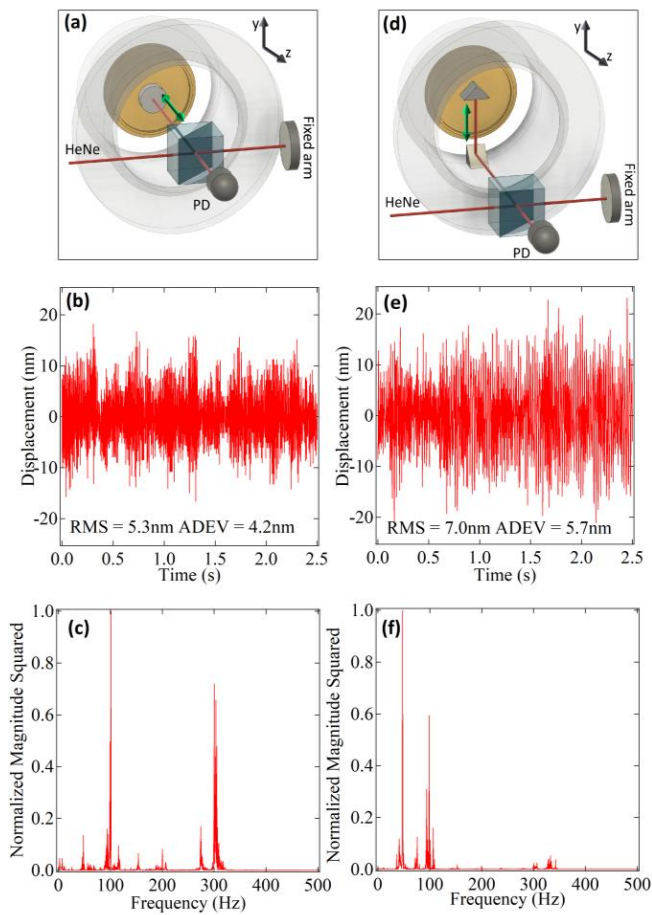

FIG. 5. Interferometric evaluation of the vibrations in the z and y directions. The sample holder is depicted as the brass colored cylindrical base at the bottom of the sample well. (a) Michelson interferometer used for the evaluation of the vibrations in the z direction. The green arrow shows the sensitive measurement axis. (b) Time trace of the sample holder displacement in z direction. (c) Normalized magnitude squared of the Fourier Transform of the time trace in (b). (d) Michelson interferometer for the measurement of the vibrations in the y direction. The bottom silver coated prism is directly attached to the sample well (cryostat casing). The green arrow shows the sensitive measurement axis. (e) Time trace of the sample holder displacement in the y direction and (f) Normalized magnitude squared of the FFT of the time trace in (e).

We also performed a similar measurement in the y-direction (vertically with respect to the optical axis of the cryostat) in a similar manner. We implemented a slightly modified Michelson interferometer sensitive to vibrations in the y-direction, like the one depicted in Fig. 5 (d). Again, the green arrow shows the sensitive measurement axis. The y-direction vibrations show slightly higher amplitude, reaching peak to peak value of ~40 nm, ~7 nm RMS and ~6 nm ADEV vibrations, as shown in Fig. 5 (e). In the y-direction, the vibrational spectra show two dominant frequencies at 47 Hz and 98 Hz, as shown in Fig. 5 (f). Both z- and y-direction measurements were taken at cryogenic temperatures of approximately 8 K on the sample holder. Comparative measurements were also taken with the stinger off at cryogenic temperatures as well, which reveals the baseline of the vibrations for our system and the lab. The values here were measured to be ~4 nm peak to peak, ~1.4 nm RMS and ~1.1 nm ADEV.

## VI. SUPERCONDUCTING MAGNET OPERATION

Energization of the superconducting magnet requires that the temperature of the sidearm (containing the high current leads) and magnet be below 35 K and 5.5 K respectively. Reaching temperatures below these values is naturally occurring with sufficient helium flow through the system. Care must be taken when energizing the magnet as the high current passing through the leads causes a temperature increase that must be factored in to avoid the risk of quenching. Fig. 6, shows the temperature behavior during a ramp up (a) and a ramp down (b) of the magnetic field. During magnet energization, shown here in Fig. 6 (a), the temperature of the sidearm initially drops by ∼2 K and then increases by ∼5 K. This phenomenon is due to evaporation of liquefied helium in the cryostat because of the heat introduced during energization of the magnet. This leads to increased helium flow rates, momentarily causing a drop in temperature. As the current further increases, the heat introduced eventually leads to the observed temperature rise. The maximum rate of change of the field we attempted during energization was 0.5 T/min leading to a 10 minute ramp up time. For this rate, the sidearm requires approximately 25 minutes before reaching a steady state. It is therefore important to ensure that the initial value of the sidearm temperature is well below the upper bound for the correct operation of the system. De-energizing the superconducting magnet can be performed at the maximum permissible rate of 1 T/min during which, the sidearm temperature experiences a temperature reduction while the magnet temperature rises to 5.2 K, as shown in Fig. 6 (b). Once the field has been brought to zero, it takes approximately 10 minutes before the system will reach a steady state. It is interesting to point out that the temperature of the cryostat exhibits small oscillations on the order of 0.25 K peak to peak. These are most visible on the sample temperature $T_{sample}$. This is the result of a complex interplay between the formation of liquid helium in the cryostat and its impact on the flowrates of helium through the system. In particular, as liquid helium is formed the pressure of the helium reduces (order of ~0.5 psi) leading to a slight warm up. This causes a small amount of helium to boil off, slightly increasing the pressure and thus the flow rates, leading to a temperature reduction and reliquefaction of helium. The repetition of this cycle manifests itself as the observed temperature oscillations.

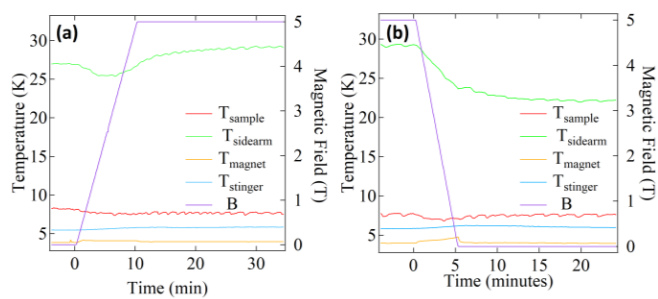

**FIG. 6**. Temperature profiles of the cryostat components and the Stinger during operation of the superconducting magnet. (a) Energization of the superconducting magnet at a magnetic field rate of 0.5 T/min. (b) De-energization of the superconducting magnet at maximal field ramping rate of 1 T/min. At this rate the magnet reaches 5.2K which is still below the limit value of 5.5K.

## VII. SUPERCONDUCTING MAGNET QUENCHING

For the magnetic cryostat we investigate here, loss of superconductivity in the magnet while it is energized, usually occurs when the magnet is energized for the first time (break in of the magnet), or by increasing the temperature of any part of the magnet beyond the critical temperature for superconductivity. When this happens, the magnet undergoes a so-called quench and the energy stored in the magnet is rapidly dissipated as heat. Fig. 7, shows one such occurrence during magnet testing. Here, the magnet quench occurred at a field strength of 4.96 T out of a maximum of 5 T. As can be seen in Fig. 7 (a), all elements of the system experience a sharp spike in temperature, which rapidly rises to approximately 40 K. The system however recovers within 40 - 60 minutes after which, it is at low enough temperatures to allow for magnet re-energization. Fig. 7 (b), shows the temporal temperature gradients for the Stinger and the 3 parts of the cryostat during the magnet quench. The temperature gradients show a big spike with the highest rate recorded for the magnet, reaching ~650 K/min. The thermal shock causes any liquefied helium to boil off, leading to a pressure spike in the manifold. A small amount of helium might be vented to the lab through the 20psi overpressure valve within the manifold. If venting does occur, topping-up the helium circuit with helium gas from a compressed helium cylinder might be required, in order to reach the operational pressure of ~100 psi.

## VIII. CONCLUSION

We have demonstrated successful operation of a continuous flow, liquid helium magnetic cryostat as a closed cycle system using a modular cryocooler. We have investigated the performance of the combined system and found stable operation and low vibrations, making this system attractive for single quantum emitter magnetic cryo-microscopy. We hope that this work will serve as a guide for research groups that wish to operate this type of magnetic cryostat in closed cycle mode.

## SUPPLEMENTARY MATERIAL

See supplementary material for additional details on the fabricated He manifold and the interferometric measurements.

## ACKNOWLEDGEMENTS

The authors thank ColdEdge Technologies for providing a demo unit of the Stinger cryocooler and Prof. P. G. Lagoudakis from the University of Southampton for providing the Oxford MicrostatMO at the Laboratories for Hybrid Optoelectronics where the initial proof of concept testing of the combined system took place.

## DATA AVAILABILITY

The data that support the findings of this study are available from the corresponding authors upon reasonable request.

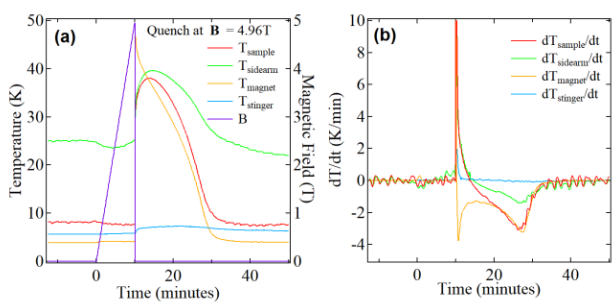

**FIG. 7.** (a) Typical temperature profiles during a magnet quench. Most components of the system experience a sharp increase in temperature within a short time. Steady state is reached within approximately 40-60 minutes before operation of the magnet can be done again. (b) Temporal temperature gradients of the cryostat components and the Stinger during the quench. The maximum recorded rate is ~650 K/min (not shown)

# Supplementary Material for

# Operation of a Continuous Flow Liquid Helium Magnetic Microscopy Cryostat as a Closed Cycle System

K. Barr[1,a)], T. Cookson[2,3] and K. G. Lagoudakis[1,a)]

## AFFILIATIONS

[1]Department of Physics, University of Strathclyde, Glasgow G4 0NG, United Kingdom
[2]Department of Physics and Astronomy, University of Southampton, Southampton SO17 1BJ, United Kingdom
[3]Centre for Photonics and Quantum Materials, Skolkovo institute of science and technology, Skolkovo, Moscow 121205, Russian Federation

[a)]Author to whom correspondence should be addressed: k.barr@strath.ac.uk and k.lagoudakis@strath.ac.uk

## ABSTRACT

We provide additional details on the microscope objective, the fabricated He manifold, the He pressure behavior during cool down and the analysis of the interferometric vibrational data.

## I. MICROSCOPE OBJECTIVE FOR MAGNETIC MICROSCOPY WITH THE MICROSTAT-MO

The magnet bore is 40mm in diameter allowing for movement of a room temperature microscope objective in order to investigate different real space locations on the sample surface. A microscope objective that we have used successfully is the Mitutoyo M Plan Apo NIR 100x (part number 378-826-15) with a 0.5 numerical aperture. This objective has a working distance of 12 mm, a max diameter of 32 mm and it does not contain ferromagnetic components making ideal for magneto-microscopy measurements. Note that the newest generation of the same objectives (part number 378-826-5) contains ferromagnetic components which might lead to significant forces applied on the objective when the magnet is operated, hence not advised for this type of experiments.

## II. CUSTOM HELIUM MANIFOLD

The manifold is housed in a 19" rack steel enclosure with 266mm height (6U height) and 300 mm depth as shown in Fig. S1. All helium line connections to and from the manifold are fitted with Eaton Aeroquip low inclusion fittings (model number 5400-S2-4) to match the connectors of the Stinger supply and return flex tubes and to minimize contamination of the helium circuit with air upon connection of the helium flex lines. The vacuum pump port was fitted with a KF-16 vacuum flange. For the pressure gauges we used glycerin filled gauges but dry gauges are suitable too as the pressures in the system are not jittering.

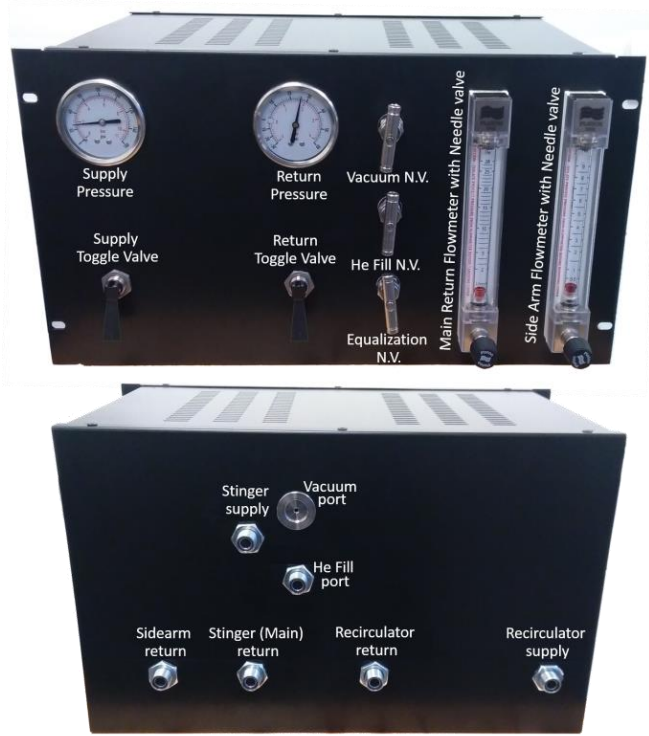

**FIG. S1**. Top: Front panel of the custom built manifold for the operation of the cryostat. Bottom: Back panel of the custom built manifold. N.V.: Needle Valve

## II. HELIUM LIQUEFACTION AND IMPACT ON PRESSURE

There are two stages during the cooling cycle at which significant changes of the He pressure occur that should be accounted for. Starting with a pressure on the order of 100-110 psi ensures that after the expected pressure drops, the pressure is high enough to allow for sufficient He flow through the system and maintain cryogenic temperatures. The first pressure drop occurs when the Stinger reaches its base temperature which at the location of the temperature probe, it is approximately 6K, around 1.5 hours into the cooling cycle. At this stage small amounts of He within the Stinger itself liquefy, leading to a small pressure drop of ~8 psi as seen in Fig. S2. After this, the pressure remains relatively constant until the 10 hour mark, where we see the bulk of liquefaction occurring. This is around the time the magnet reaches its base temperature of approximately 4K and liquefaction occurs inside the cryostat. This leads to the pressure dropping an additional ~40 psi. Should the starting pressure be too low, this liquefaction may prevent sufficient helium flows through the cryostat and cause the system to be unable to either maintain or reach the temperatures sought after.

## III. POST-PROCESSING OF INTERFEROMETRIC DATA

The main paper showed post-processed interferometric data that allowed one to directly read off the peak to peak displacement amplitudes due to vibrations, as well as allowing for the calculation of the RMS and ADEV of the vibrations. A frequency stabilized HeNe laser was used ($\lambda$=632.8 nm) to setup both the Michelson interferometers with readout occurring through a fast photodiode. Externally perturbing the system allows one to collect the peak to peak signal of the interference, on the photodiode. The recorded voltage is directly related to the fringe maximum and fringe minimum of the interferometer corresponding to $\lambda/4$ in nanometers. Peak to peak data (V) allowed for evaluation of the nanometers per volt (nm/V) at which point the data was rescaled to show displacement in nanometers as seen in Fig. S3. From here, smoothing of the signal using a rolling window averaging technique was performed allowing one to evaluate the slow varying drift of the signal whilst not removing any of the signal due to vibrations. This slow variation could then be subtracted element wise from the original data array to produce a vibrational signal that is centered around zero.

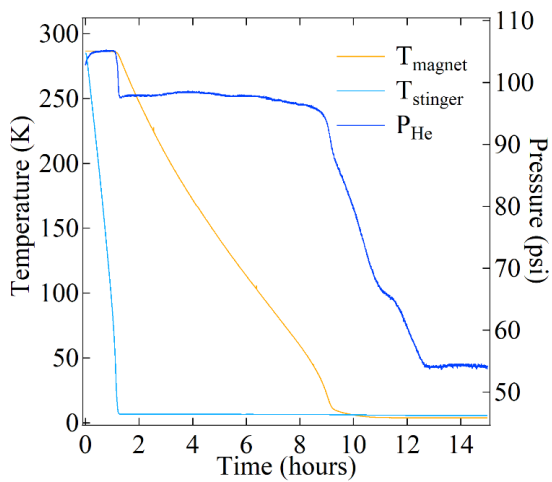

**FIG. S2**. Temperature of magnet and Stinger on a cool down cycle (left) with helium pressure (right). First pressure drop occurs when the Stinger reaches ~6K (1.5 hours), and the second continuous pressure drop happens when the magnet reaches ~4K (9.5 hours).

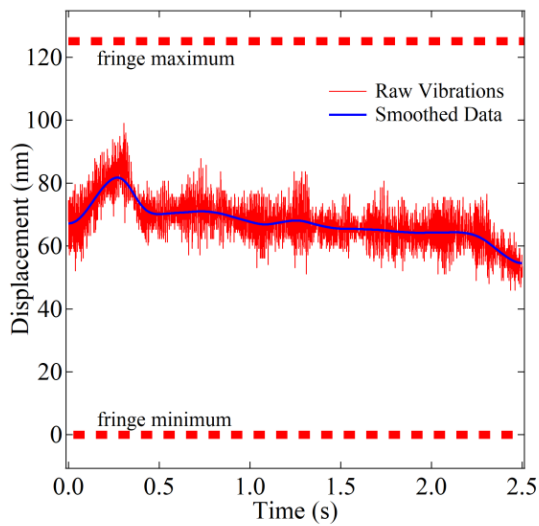

**FIG. S3**. Displacement of signal due to vibrations (red) and slow drift (blue). Data is always acquired when the signal is within the 30-70% band the fringe amplitude where the system is most sensitive. The slow drift is produced by a rolling average of the data and does not remove any of the vibrations but rather permits to center the data with zero mean.